\documentclass[conference,10pt]{IEEEtran}
\usepackage{epsfig}
\usepackage{indentfirst}
\usepackage{amssymb}
\usepackage{cite}      % Written by Donald Arseneau
\usepackage{graphicx}
\usepackage{psfrag}    % Written by Craig Barratt, Michael C. Grant,
\usepackage{subfigure} % Written by Steven Douglas Cochran
\usepackage{url}       % Written by Donald Arseneau
\usepackage{amsmath}   % From the American Mathematical Society
\usepackage{array}
\usepackage[hypertex]{hyperref}

\begin{document}

% paper title
\title{Revisiting the Issues On Netflow Sample and Export Performance}

% author names and affiliations
% use a multiple column layout for up to three different
% affiliations
\author{\authorblockN{Hamed Haddadi, Raul Landa, Miguel Rio}
\authorblockA{Department of Electronic and Electrical Engineering\\
University College London\\
United Kingdom\\
Email: hamed,mrio,rlanda@ee.ucl.ac.uk}
\and
\authorblockN{Saleem Bhatti}
\authorblockA{School of Computer Science\\
University of St. Andrews\\
United Kingdom\\
Email: saleem@dcs.st-and.ac.uk}}

% make the title area
\maketitle

\begin{abstract}
\emph{The high volume of packets and packet rates of traffic on some router links makes it exceedingly difficult for routers to examine every packet in order to keep detailed statistics about the traffic which is traversing the router. Sampling is commonly applied on routers in order to limit the load incurred by the collection of information that the router has to undertake when evaluating flow information for monitoring purposes. The sampling process in nearly all cases is a deterministic process of choosing 1 in every $N$ packets on a per-interface basis, and then forming the flow  statistics based on the collected sampled statistics. Even though this sampling may not be significant for some statistics, such as packet rate, others can be severely distorted. However, it is important to consider the sampling techniques and their relative accuracy when applied to different traffic patterns. \\The main disadvantage of sampling is the loss of accuracy in the collected trace when compared to the original traffic stream. To date there has not been a detailed analysis of the impact of sampling at a router in various traffic profiles and flow criteria. In this paper, we assess the performance of the sampling process as used in NetFlow in detail, and we discuss some techniques for the compensation of loss of monitoring detail.}

\end{abstract}

% no keywords

% For peer review papers, you can put extra information on the cover
% page as needed:
% \begin{center} \bfseries EDICS Category: 3-BBND \end{center}
%
% for peerreview papers, inserts a page break and creates the second title.
% Will be ignored for other modes.
\IEEEpeerreviewmaketitle

\section{Introduction}

Packet sampling is an integral part of passive network measurement on today's Internet. The high traffic volumes on backbone networks and the pressure on routers has resulted in the need to control the consumption of resources in the measurement infrastructure. This has resulted in the definition and use of estimated statistics by routers, generated based on sampling packets in each direction of each port on the routers. The aims of this paper is to analyse the effects of the sampling process as operated by NetFlow, the dominant standard on today's routers.

There are three constraints on a core router which lead to the use packet sampling: the size of the record buffer, the CPU speed and the record look-up time. In \cite{ciscobook}, it is noted that in order to manage and analyse the performance of a network, it is enough to look at the basic statistical measures and summary statistics such as average range, variance, and standard deviation. However, in this paper we analyse both analytically and practically the accuracy of the inference of original characteristics from the sampled stream when higher order statistics are used.

This paper focuses on the inference of original network traffic characteristics for flows from a sampled set of packets and examines how the sampling process can affect the quality of the results. In this context, a flow is identified specifically, as the tuple of the following five key fields: Source IP address, Destination IP address, Source port number, Destination port number, Layer 4 protocol type.

\subsection{NetFlow memory constraints}

A router at the core of an internet link is carrying a large number of flows at any given time. this pressure on the router entails the use of strict rules in order to export the statistics and keep the router memory buffer and CPU resources available to deal with changes in traffic patterns by avoiding the handling of large tables of flow records. Rules for expiring NetFlow cache entries include:

\begin{itemize}

\item Flows which have been idle for a specified time are expired and removed from the cache (15 seconds is default)

\item Long lived flows are expired and removed from the cache (30 minutes is default)

\item As the cache becomes full a number of heuristics are applied to aggressively age groups of flows simultaneously

\item TCP connections which have reached the end of byte stream (FIN) or which have been reset (RST) will be expired

\end{itemize}

\subsection{Sampling basics}

Distributions studies have been done extensively in literature. In brief conclusion, internet traffic is believed to have Heavy-tailed distribution, self-similar nature, Long Range Dependence \cite{selfsim}. Sampling has the following effects on the flows:

\begin{itemize}
\item It is easy to miss short flows \cite{nickrev}
\item Mis-ranking on high flows \cite{chadi}
\item Sparse flow creation\cite{nickrev}\end{itemize}

\emph{Packet sampling:}\\
The inversion methods are of little to no use in practice for low sampling probability $q$, such as $q = 0.01$ (1 packet in 100) or smaller, and become much worse as $q$ becomes smaller still. For example, on the Abilene network, 50\% sampling was needed to detect the top flow correctly \cite{chadi}.

\emph{Flow sampling:}\\
Preserves flows intact and the sampling is done on the flow records. In practice, any attempt to gather flow statistics involves classifying individual packets into flows. All packet meta-data has to be organised into flows before sampling can take place. This involves more CPU load and more memory if one uses the traditional hash table approach with one entry per flow. New flow classification techniques, such as bitmap algorithms, could be applied but there is no practical usage in this manner currently.

\section{Variation of Higher Order statistics}

In this section we look at a more detailed analysis of the effect of sampling as performed by netflow on higher order statistics of the packet and flow size distributions. For the analysis of packet sampling application is used by NetFlow, we emulated the NetFlow operation on a 1 hour OC-48 trace, collected from the CAIDA link on $24^{th}$ of April 2003. This data set is available from the public repository at CAIDA \cite{caida}. The trace comprises of 84579462 packets with anonymised source and destination IP addresses. An important factor to rememberer in this work is the fact that the memory constraint on the router has been relaxed in generating the flows from the sampled stream. This means that there maybe more than tens of thousands of flow keys present at the memory at a given time, while in NetFlow, the export mechanism empties the buffer list regularly which can have a more severe impact on the resultant distribution of flow rates and statistics\footnote[3]{The processing of the data was done using tools which are made available to the public by the authors.}.

\subsection{Effects of the short time-out imposed by memory constraints}

Table \ref{tabledt} illustrates the data rates $d(t)$ per interval of measurement. Inverted data rates, by dividing $d(t)$ by the sampling probability $q$, are shown as $dn(t)$.

\begin{table}[hbtp]
\centering
\caption{The statistical properties on Data rates $d(t)$}
\begin{tabular}{||c|c|c|c|c||} \hline \hline
Dataset,bin(secs)		& STD		& Skewness 	& Kurtosis	\\ \hline \hline
$d(t)$, 30		& 2.2274e+07	& 0.5421  	& 0.6163 	\\ 
$d_{n}(t)$, 30		& 2.9109e+07	& 0.3837	& 0.4444 	\\ \hline \hline
$d(n) - d_{n}(t)$, {\em 30}		& {\em 1.6748e+07}	& {\em -0.2083}	& {\em 0.7172}	\\ \hline \hline
$d(t)$, 120		& 7.8650e+07	& 0.7398	& 1.6190\\ 
$dn(t)$, 120		& 9.5216e+07	& 0.3274	& 0.9268\\ \hline \hline
$d(t)-d_{n}(t)$, {\em 120}		& {\em 3.7652e+07}	& {\em -0.2971}	& {\em -1.1848}	\\ \hline \hline
$d(t)$, 300		& 1.8491e+08	& 1.3058	& 3.7451\\ 
$d_{n}(t)$, 300		& 2.1248e+08	& 1.1016	& 2.5408\\ \hline \hline
$d(t)-d_{n}(t)$, {\em 300}		& {\em 6.1039e+07}	& {\em 0.1840}	& {\em -1.1628}	\\ \hline \hline
\end{tabular}
\label{tabledt}
\end{table}

As observed in table \ref{tabledt}, the mean does not have a great variation, possibly because distributions of packet sizes within single flows do not exhibit high variability. The standard deviation of the estimated data rate is higher than the corresponding standard deviation for the unsampled data stream. In the absence of any  additional knowledge about the higher level protocol, or the nature  of the session level activity, in the unsampled data stream, each flow can be thought of as having packets of varying sizes that are more or less independent from one another. Thus, the whole traffic profile results from the addition of many independent random variables which, by the central limit theorem, tend to balance among themselves to produce a more predictable, homogeneous traffic aggregate. However, simple inversion eliminates this multiplicity of randomly distributed values by introducing a very strong correlation effect, whereby the size of all the packets in a reconstructed flow depend on the size of a very small set of sampled packets. This eliminates the possibility for balancing and thus increases the variability of the resulting stream, i.e. its standard deviation. 

However, the skewness and kurtosis do change. Skewness is a measure of the asymmetry of the probability distribution of a real-valued random variable. Roughly speaking, a distribution has positive skew (right-skewed) if the right (higher value) tail is longer and negative skew (left-skewed) if the left (lower value) tail is longer (confusing the two is a common error). Skewness, the third standardised moment, is written as $\gamma_1$ and defined as:
\begin{center}
$\gamma_1 = \frac{\mu_3}{\sigma^3}$
\end{center}

where $\mu_{3}$ is the third moment about the mean and $\sigma$ is the standard deviation.

Kurtosis is more commonly defined as the fourth cumulant divided by the square of the variance of the probability distribution,

\begin{center}
$\gamma_2 = \frac{\kappa_4}{\kappa_2^2} = \frac{\mu_4}{\sigma^4} - 3$
\end{center}

which is known as excess kurtosis. The "minus 3" at the end of this formula is often explained as a correction to make the kurtosis of the normal distribution equal to zero. The skewness is a sort of measure of the asymmetry of the distribution function. The kurtosis measures the flatness of the distribution function compared to what would be expected from a Gaussian distribution. 

Table \ref{tablept} illustrates the packet rates $p(t)$ per interval of measurement. Inverted packet rates, by dividing $p(t)$ by the sampling probability $q$, are shown as $pn(t)$. The distributions before and after sampling are extremely close, and thus their difference tends to exaggerate those small difference that they do have. That is the reason of the enormous skewness and kurtosis that are observed. The skewness of the reconstructed stream is smaller than that of the unsampled stream this means that the reconstructed distribution is more symmetric, that is , it tends to diverge in a more homogeneous manner around the mean. Additionally, it is positive, meaning that in both cases the distribution tends to have longer tails towards large packets rather than towards short packets, concentrating its bulk on the smaller packets. If we concede that small flows (flows consisting of a small number of packets) tend to contain small packets, then it is clear that this smaller packets will be underrepresented and the distribution will shift its weight towards bigger packets (members of bigger flows). Thus, it will become more symmetric and hence less skewed.

\begin{table}[hbtp]
\centering
\caption{The statistical properties on Packet rates $p(t)$}
\begin{tabular}{||c|c|c|c|c||} \hline \hline
Dataset,bin(secs)		& STD		& Skewness 	& Kurtosis	\\ \hline \hline
$p(t)$, 30		& 3.1162e+04	& -0.4007	& 0.7415  \\  
$p_{n}(t)$, 30		& 3.1359e+04	& -0.3584	& 0.6072\\ \hline \hline
$p(t)-p_{n}(t)$, {\em 30}		& {\em 5.4148e+03}	& {\em 9.1469}	& {\em 96.0659}	\\ \hline \hline
p(t), 120		& 1.1215e+05	& -0.3875	& 1.2027\\ 
pn(t), 120		& 1.1178e+05	& -0.3759	& 1.2238\\ \hline \hline
$p(t)-p_{n}(t)$, {\em 120}		& {\em 3.0157e+03}	& {\em 4.7140}	& {\em 26.1079}\\ \hline \hline
p(t), 300		& 2.5128e+05	& 0.1305	& 1.6495\\ 
pn(t), 300		& 2.5152e+05	& 0.1433	& 1.6597\\ \hline \hline
$p(t)-p_{n}(t)$, {\em 300}		& {\em 2.1047e+03}	& {\em 2.4298}	& {\em 8.9377}\\ \hline \hline
\end{tabular}
\label{tablept}
\end{table}

The Kurtosis decreases in all of the considered examples. This means that the reconstructed streams are more homogeneous and less prone to outliers when compared with the original traces. Thus, more of the variance in the original traces in packet size can be attributed to infrequent packets that have inordinately big packets that were missed in the sampling process, and thus the variance in the reconstructed stream consists more of homogeneous differences and not large outliers. However, both the reconstructed and unsampled streams are leptokurtic and thus tend to have long, heavy tails.

\subsection{The two-sample KS test}

The two-sample KS test is one of the most useful and general non-parametric methods for comparing two samples, as it is sensitive to differences in both location and shape of the empirical cumulative distribution functions of the two samples. A CDF was calculated for the number of packets per flow and the number of octets per flow for each of the 120 sampling intervals of 30 seconds each, both for the sampled/inverted and unsampled streams. Then, a Two-Sample Kolmogorov-Smirnov Test with $5\%$ significance level was performed between the 120 unsampled and the 120 sampled \& inverted distributions. In every case the distributions before and after sampling and inversion were found to be significantly different, and thus it is very clear that the sampling and inversion
process significantly distorts the actual flow behaviour of the network.

\section{Practical Implications of Sampling}

The effects of sampling on network traffic statistics can be measured from different perspectives. In this section we will cover the theories behind the sampling strategy and use some real data captures from CAIDA in an emulation approach to demonstrate the performance constraints of systematic sampling.

\subsection{Inversion errors on sampled statistics}

\begin{figure}[htbp]
	\centering
	\includegraphics[width=4in,height=2in,bb=   58   215   546   602]{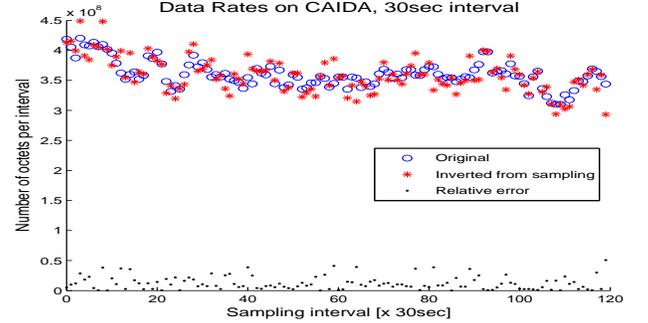}
% nosamp.eps: 300dpi, width=4.13cm, height=3.28cm, bb=   58   215   546   602
	\caption{Data rates per 30 second interval, original versus normal inversion of sampled}
\label{compare}
\end{figure}

The great advantage of sampling is the fact that the first order statistics do not show much variation when the sampling is done at consistent intervals and from a large pool of data. This enables the network monitoring to use the sampled statistics to form a relatively good measure of the aggregate measure of network performance. Figure \ref{compare} displays the data rates $d(t)$, in number of bytes seen per 30 second interval, on the one hour trace. The inverted data $d(t)$ is also shown with diamond notation, showing the statistics gathered after the sampled data is multiplied by the sampling rate. The black dots display the relative error per interval, $e(t)= \frac{d(t) - dn(t)}{d(t)}$.

\begin{figure}[htbp]
	\centering
	\includegraphics[width=3in,height=1.5in,bb=   58   215   546   602]{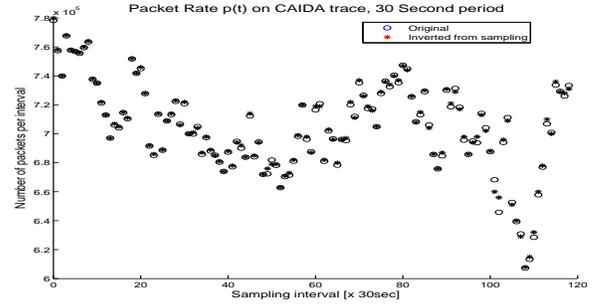}
% nosamp.eps: 300dpi, width=4.13cm, height=3.28cm, bb=   58   215   546   602
	\caption{Packet rates per 30 second interval, original vs inversion of sampled}
\label{pktscompare}
\end{figure}

Figure \ref{pktscompare} displays the packet rates $p(t)$, the number of packets per 30 second interval, versus the sampled and inverted packet rates $pn(t)$. In this figure, it can be observed that the inversion does a very good job at nearly all times and the relative error is negligible. This is a characteristics of systematic sampling and is due to the central limit theorem.

It can be readily seen that the recovery of packet rates by simple inversion is much better than the recovery of data rates. This is because sampling one in a thousand packets deterministically can be trivially inverted by multiplying by the sampling rate (1000): we focus on packet level measurement, as opposed to a flow level measurement. If the whole traffic flow is collapsed into a single link, then if we sample one packet out every thousand and then multiply that by the sampling rate, we will get the total number of packets in that time window. We believe that the small differences that we can see in Figure \ref{pktscompare} are due to the fact that at the end of the window some packets are lost (because their `representative' was not sampled) or overcounted (a `representative' for 1000 packets was sampled but the time interval finished before they had passed). We believe these errors happen between measurement windows in time, i.e. they are window-edge effects.

\begin{figure}[htp]
	\centering
	\includegraphics[width=3in,height=1.5in,bb=   58   215   546   602]{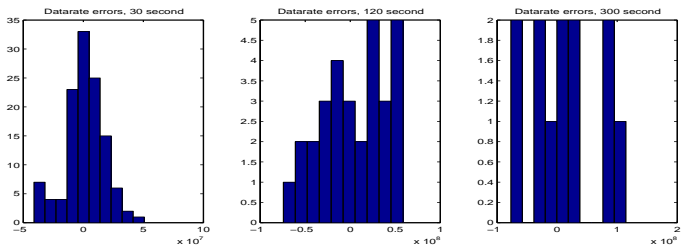}
% nosamp.eps: 300dpi, width=4.13cm, height=3.28cm, bb=   58   215   546   602
	\caption{Standard Sampling \& inversion error on data rates, different measurement bins}
\label{dataerr}
\end{figure}

\begin{figure}[htp]
	\centering
	\includegraphics[width=3in,height=1.3in,bb=   58   215   546   602]{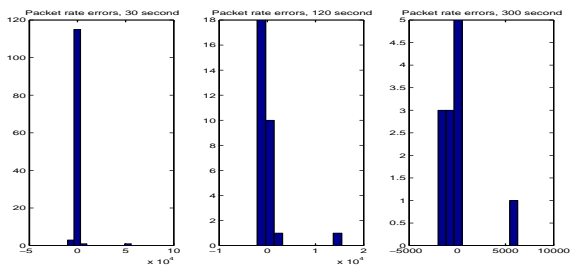}
% nosamp.eps: 300dpi, width=4.13cm, height=3.28cm, bb=   58   215   546   602
	\caption{Sampling \& inversion error on packet rates, different measurement bins}
\label{pkterror}
\end{figure}

The inversion property described above does not hold for measuring the number of bytes in a sampling interval. Simple inversion essentially assumes that all packets in a given flow are the same size, and of course this assumption is incorrect. It is to be expected that the greater the standard deviation of packet size over an individual flow, the more inaccurate the recovery by simple inversion will be regarding the number of bytes per measurement interval. Figures \ref{dataerr} and \ref{pkterror} displays the standard error rate on data rate and packet rate recovery respectively, in different measurement intervals.

\subsection{Flow size and packet size distributions}

\begin{figure}[htp]
	\centering
	\includegraphics[width=3in,height=1.5in,bb=   58   215   546   602]{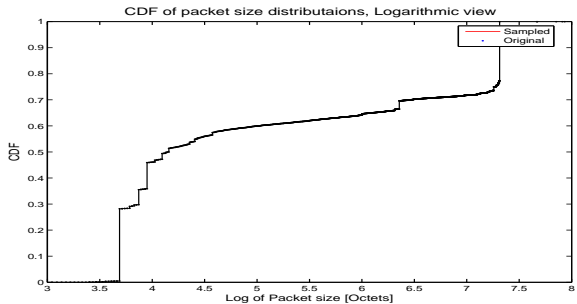}
% nosamp.eps: 300dpi, width=4.13cm, height=3.28cm, bb=   58   215   546   602
	\caption{Normalised CDF of packets distributions per flow, original vs inverted}
\label{pcktscdf}
\end{figure}

Figure \ref{pcktscdf}, displays the CDF of packet size distribution in all the flows formed from the sampled and unsampled streams. The little variation in the packet size distribution conforms to the findings of the previous section where it was discussed that the packet sampling has low impact on the packet size distribution.

\begin{figure}[htp]
\centering
\includegraphics[width=0.45\textwidth]{{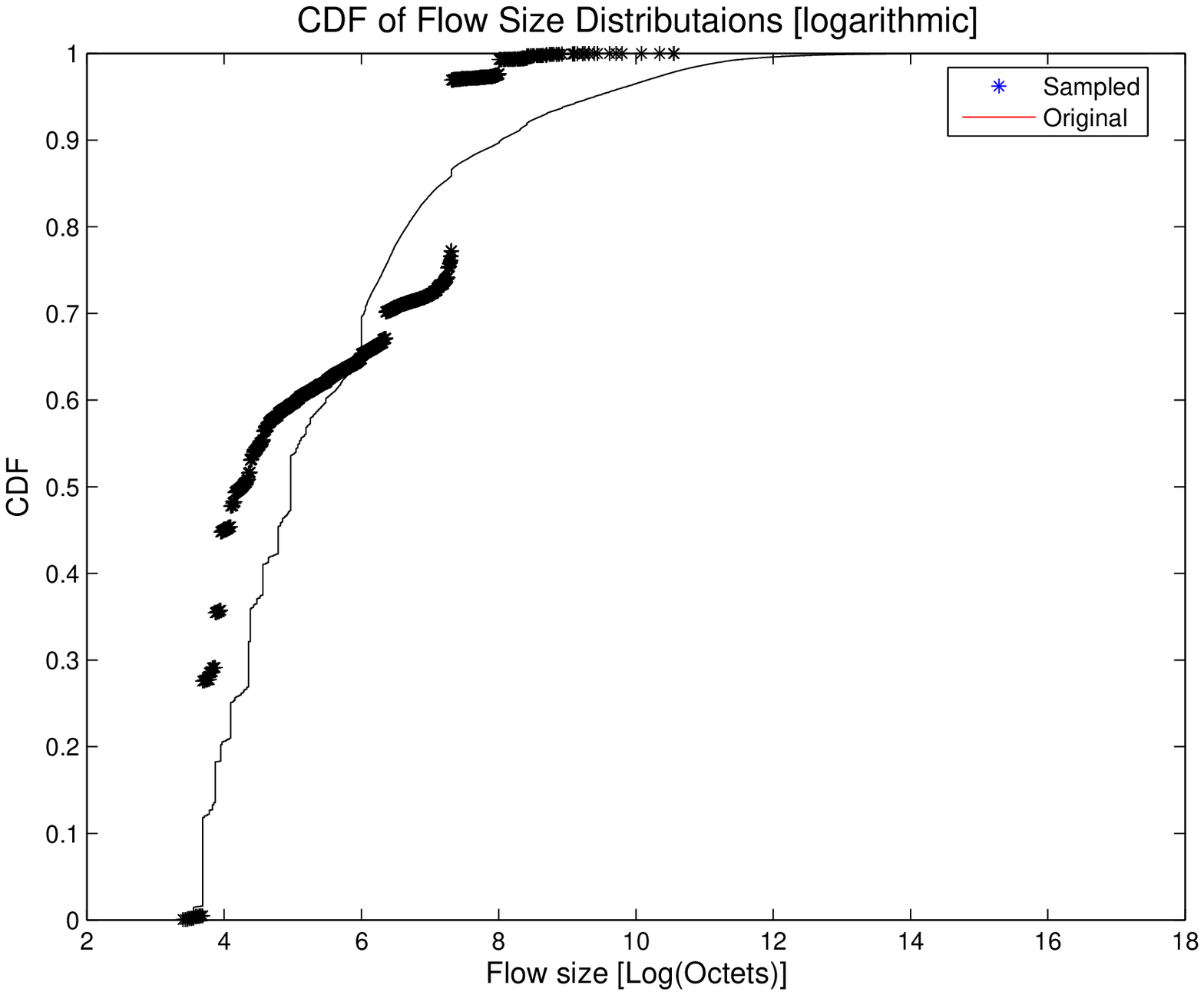}}
\hfill
\includegraphics[width=0.45\textwidth]{{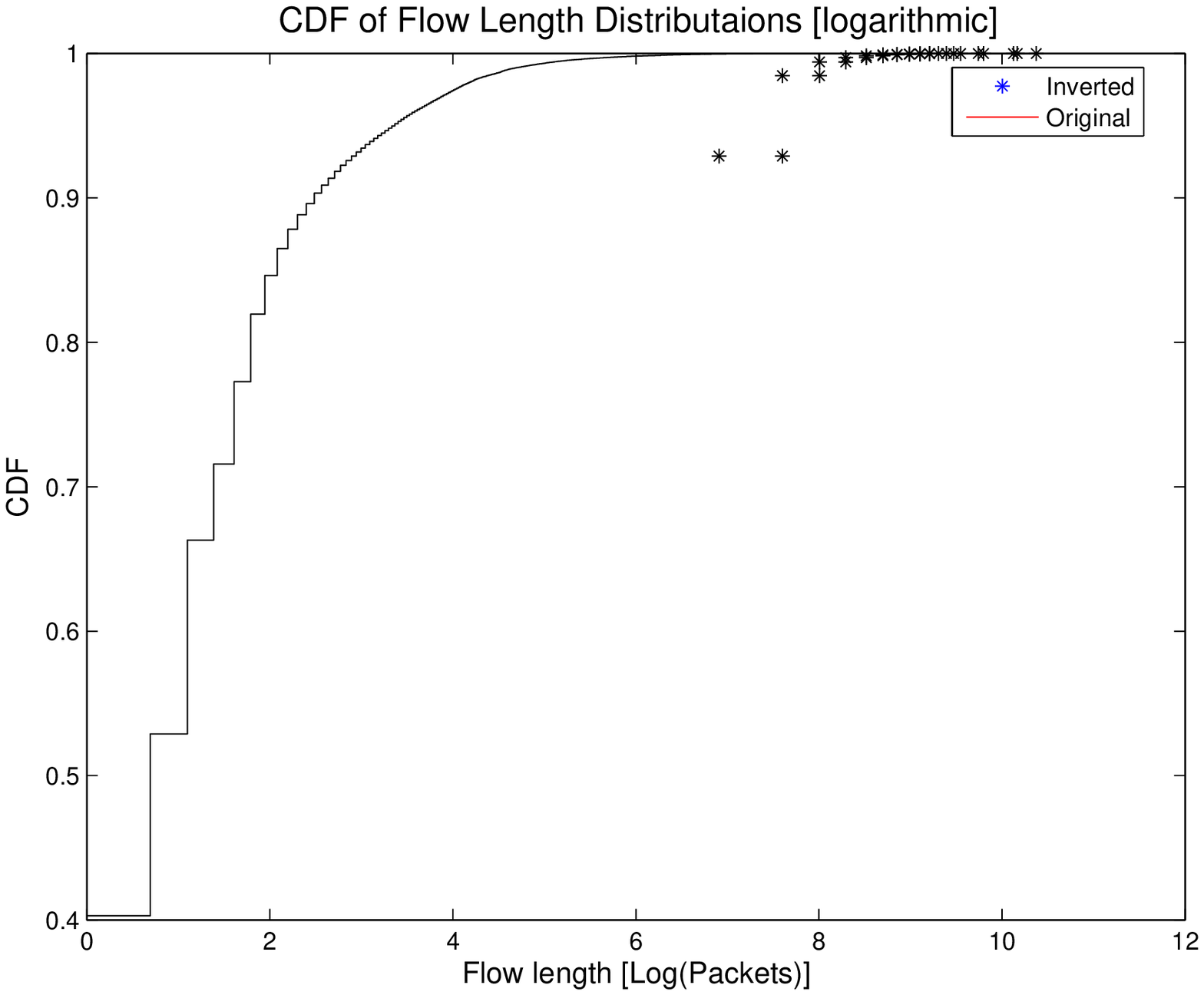}}
\caption{Normalised CDF of flow size in packets [figure] \& length in bytes [right] per flow, original vs inverted}
\label{flowpckts}
\end{figure}

Figure \ref{flowpckts}:1 shows the effect that the distribution of packet lengths can have on the distribution of flow lengths when periodic packet sampling is applied. As flows reconstructed from a sampled packet stream are predominantly formed by just one packet, their length distribution follows that of single packets (Figure \ref{pcktscdf}). That is the reason for the sharp jump near 1500 octets, as this characteristic originates from the maximum frame size in ethernet networks.

% \begin{figure}[htp]
% 	\centering
% 	\includegraphics[width=3.5in,height=1.5in,bb=   58   215   546   602]{images/flowsizeCDFlog.eps}
% % nosamp.eps: 300dpi, width=4.13cm, height=3.28cm, bb=   58   215   546   602
% 	\caption{Normalised CDF of flow size in packets per flow, original vs inverted}
% \label{flowscdf}
% \end{figure}

From Figure \ref{flowpckts}:2 , it can be readily seen that, in the sampled stream, more than 90 percent of flows consist of a single packet, whereas in the unsampled case a much grater diversity in flow lengths exists for small flows. This is due to the fact that simple packet-based deterministic sampling under-represents short flows, and those short flows that are indeed detected by the procedure after sampling usually consist of a single packet. Thus, short flows are either lost or recovered as single packet flows, and long flows have their lengths reduced.

% \begin{figure}[htp]
% 	\centering
% 	\includegraphics[width=3.5in,height=1.5in,bb=   58   215   546   602]{images/flowlengthCDFloginverted.eps}
% nosamp.eps: 300dpi, width=4.13cm, height=3.28cm, bb=   58   215   546   602
% 	\caption{Normalised CDF, number of packets per flow, original vs inverted}
% \label{flowpckts}
% \end{figure}

\section{Conclusion}

In this paper we have reviewed the effects of sampling and flow record creation, as done by NetFlow, on the traffic statistics which are reported by such a process. It is inevitable that systematic sampling can no longer provide a realistic picture of the traffic profile present on internet links. The emergence of applications such as video on-demand, file sharing, streaming applications and even on-line data processing packages prevents the routers from reporting an optimal measure of the traffic traversing them. In the inversion process, it is a mistake to assume that the inversion of statistics by multiplication by the sampling rate is an indicate of even the first order statistics such as packet rates.

An extension to this work and the inversion problem entails the use of more detailed statistics such as port numbers and TCP flags in order to be able to infer the original characteristics from the probability distribution functions of such variables. This will enable a more detailed recovery of original packet and data rates for different applications. The inference of such probabilities, plus use of methods such as Bayesian inference, would enable a forecasting method which would enable the inversion of the sampled stream in near real time.

In a related work, we will be looking at alternative flow synthesis schemes, looking at techniques replacing the NetFlow, such as use of hashing techniques using Bloom filters. The use of a light weight flow indexing system will allow for a larger number of flows to be present at the router, possibly increasing the memory constraints and allowing for a higher sampling rate, which will in turn lead to more accurate inversion.

\section{Related Work}

There has been a great deal of worked done on analysis of sampling process and inversion problem. Choi et al. have explored the sampling error and measurement overhead of NetFlow in \cite{choi} though they have not looked at inversion process.

In \cite{loss}, the authors have compared the Netflow reports with those obtained from SNMP statistics and packet level traces, but without using the sampling feature of NetFlow which is perhaps the dominant version in use nowadays. Estan et al. \cite{betternetflow} have proposed a novel method of adapting the sampling rate at a NetFlow router in order to keep the memory resources at a constant level. This is done by upgrading the router firmware, which can be compromised by an attacker injecting varying traffic volume in order to take down the router. Also this work has not considered the flow length statistics which are the primary focus of our work.

Hohn et al. \cite{inverting} have proposed a flow sampling model which can be used in an offline analysis of flow records formed from an unsampled packet stream. In this model the statistics of the original stream are recovered to a great extent. However the intensive computing and memory resources needed in this process prevents the implementation of such a scheme on highspeed routers. They prove it impossible to accurately recover statistics from a packet sampled stream, but based on the assumption of packets being independent and identically distributed

Roughan at \cite{active} has looked at statistical processes of active measurement using Poisson and uniform sampling and has compared the theoretical performance of the two methods.  Papagiannaki et al. at \cite{dina} have discussed the effect of sampling on tiny flows when looking at generation of traffic matrices.

Authors at \cite{daniela} have been looking at anomaly detection using flow statistics, but without sampling. In \cite{nickestimate} and \cite{duffield}, authors have looked at inferring the numbers and lengths of flows of original traffic that evaded sampling altogether. They have looked at inversion via multiplication.

\section*{Acknowledgements} The authors would like to acknowledge CAIDA\cite{caida} for providing the trace files. This work is conducted under the MASTS (EPSRC grant GR/T10503) and the 46PaQ project (EPSRC grant GR/S93707).

% that's all folks
\end{document}